\documentclass[twocolumn,showpacs,preprintnumbers,amsmath,amssymb,superscriptaddress]{revtex4}

\usepackage{calc}
\usepackage{graphicx}
\usepackage{bm}

\begin{document}

\title{Optimal reaction time for surface-mediated  diffusion}


\author{O. B\'enichou}
\affiliation{Laboratoire de Physique Th\'eorique de la Mati\`ere Condens\'ee
(UMR 7600), case courrier 121, Universit\'e Paris 6, 4 Place Jussieu, 75255
Paris Cedex}

\author{D. Grebenkov}
\affiliation{Laboratoire de Physique de la Matiere Condensee (UMR 7643),
CNRS -- Ecole Polytechnique, F-91128 Palaiseau Cedex France}

\author{P. Levitz}
\affiliation{Laboratoire de Physique de la Matiere Condensee (UMR 7643),
CNRS -- Ecole Polytechnique, F-91128 Palaiseau Cedex France}

\author{C. Loverdo}
\affiliation{Laboratoire de Physique Th\'eorique de la Mati\`ere Condens\'ee
(UMR 7600), case courrier 121, Universit\'e Paris 6, 4 Place Jussieu, 75255
Paris Cedex}

\author{R. Voituriez}
\affiliation{Laboratoire de Physique Th\'eorique de la Mati\`ere Condens\'ee
(UMR 7600), case courrier 121, Universit\'e Paris 6, 4 Place Jussieu, 75255
Paris Cedex}

\date{\today}

\begin{abstract}
We present  an exact calculation of the mean first-passage time to a small target on the surface of a 2D or 3D spherical domain, for a molecule performing surface-mediated diffusion. This minimal model of interfacial reactions, which explicitly  
takes into account the combination of surface and bulk diffusion, shows the importance of correlations induced by the coupling of the switching dynamics to the geometry of the confinement, ignored so far.   Interestingly,  our results show  that, in the context of interfacial systems in confinement, the reaction time can be minimized as a function of the desorption rate from the surface, which puts forward a general mechanism of enhancement and 
regulation of chemical and biological reactivity.  
\end{abstract}


\maketitle



Among reactions limited by transport, interfacial reactions, for which molecules react on   target sites located on  the surface of the  confining domain \cite{Astumian:1985}, play an important role in situations as various as heterogeneous catalysis \cite{bond}, reactions in porous media or biochemical reactions on DNA \cite{Berg:1981} and in vesicular systems \cite{Adam:1968,Sano:1981,Schuss:2007}. Besides being a problem of great practical importance, the modeling of such systems raises two types of theoretical issues: (i) to determine the impact of geometrical parameters of the confining domain, such as the volume, on reaction kinetics, and (ii) to account for the reactive trajectories which combine bulk and surface-mediated diffusion phases due to the affinity of the molecules for the surface (see fig.1). 

 The point (i) has been studied  in the case of a perfectly reflecting  surface \cite{Ward:1993,Grigoriev:2002a,Schuss:2007,Singer:2006b,greb} and a universal scaling with the confining volume  was found  for  the mean first-passage time (MFPT) to a reactive target \cite{Sylvain_package,package_nature}. On the other hand, reactive paths of point (ii) can be described as  trajectories involving two states, one adsorbed on the surface and one desorbed in the bulk. Such two-state paths have been studied in the broader context of intermittent search strategies under the hypothesis that the times spent in each state are controlled by an internal clock independent of any geometrical parameter \cite{package_animaux,Obenichou:2008,Reingruber:2009b}. In most cases, the sojourn times in each state have been assumed to be exponentially distributed \cite{Obenichou:2008}, with the notable exception of Levy \cite{Lomholt:2008} and deterministic laws \cite{Benichou:2007,Oshanin:2007a}. 

However, in the case of interfacial reactions in confinement,   the time spent in a  bulk excursion is controlled by the statistics of return to the surface and therefore by  the geometry of the confining domain \cite{Majumdar:1999,Benichou:2005a,package_Levitz,Chechkin:2009}. Hence, this return time is not an external parameter but is generated by the very dynamics  of the diffusing molecule in confinement. As a result,  
the usual methods to calculate mean search times for  intermittent processes only provide a mean-field (MF) approximation of the reaction time which completely ignores the correlations induced by the coupling of the switching dynamics to the geometry of the confinement (see \cite{Obenichou:2008} for review, and more recently \cite{gleb}).

 Here we calculate exactly in the representative example of a spherical confining domain the MFPT to a  reactive site on the sphere for a Brownian molecule alternating phases of bulk and surface diffusion. Our analytical approach fully complies with points (i) and (ii) and shows  that 
 correlations actually strongly impact on reaction times, which are substantially underestimated by standard MF treatments.  In addition, we discuss the problem of the minimization of the reaction time with respect to the     mean adsorption time
 of the molecule on the surface, which is {\it a priori} not clear. Indeed,  the benefit of bulk diffusion, even if much faster than surface diffusion,  is questionable, since  the mean time spent in bulk excursions diverges 
with the volume of the confining domain. Surprisingly enough, we will show that, even for bulk and surface diffusion coefficients of the same order of magnitude, the reaction time can be minimized, whereas the MF treatment of \cite{gleb}
predicts a monotonic behavior. 

At the theoretical level, these results bring an exact solution to an extension to the so-called narrow escape problem (time needed to escape through a small window of an otherwise reflecting domain), which has attracted a lot of attention in last few years both in the mathematical \cite{Singer:2006b,Schuss:2007,Pillay:2010,Cheviakov:2010} and physical \cite{Benichou:2008,gleb} literature, partly due to the challenge of taking into account mixed boundary conditions. At the level of bio and chemical  physics, it puts forward a general mechanism of enhancement and regulation of  chemical reactivity by tuning the affinity of the reactants for the surface of the confining domain.

\begin{figure}[hbt]
\begin{centering}
\includegraphics[width=0.35\textwidth]{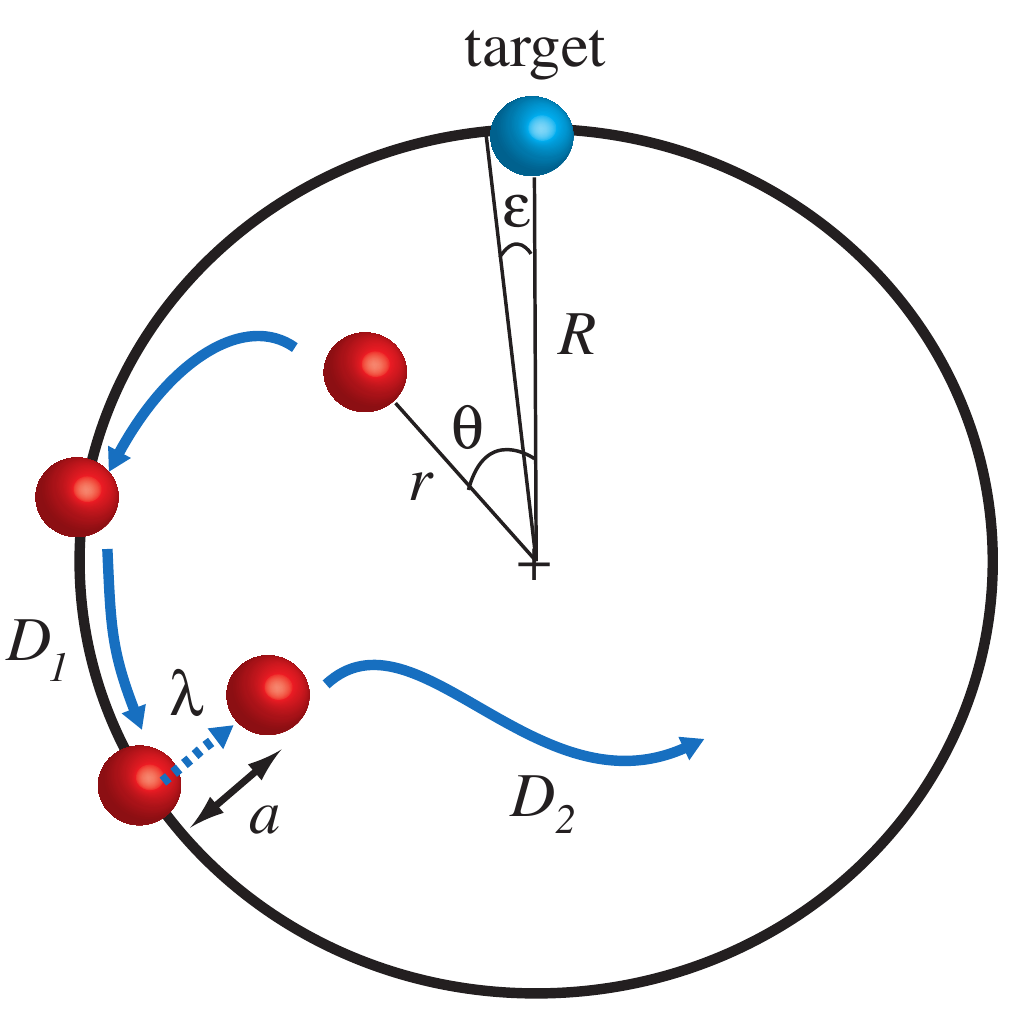}
\caption{\label{fig1}Model of surface-mediated reaction in confinement.}
\end{centering}
\end{figure}

As an archetype of confined interfacial systems, we consider a molecule in a spherical domain $S$ of radius $R$  (see fig. \ref{fig1}), alternating  phases of surface diffusion  on $\partial S$ with diffusion coefficient $D_1$    and phases of bulk diffusion in $S$ with diffusion coefficient $D_2$. The time spent during each surface phase
 is assumed to follow an exponential law with desorption
rate $\lambda$, which is reminiscent  of a first-order kinetics. At each desorption event, the molecule is assumed to be radially ejected at a distance $a$ from the surface (otherwise it is instantaneously readsorbed). Although formulated for any value of this parameter $a$ smaller than $R$, 
in most situations of physical interest $a\ll R$.  For the sake of simplicity, we first present the case of a point-like target on a 2D sphere (disk), and then generalize our approach to  targets of angular extension $2\epsilon$ on 2D and 3D  spheres.

{\it Point-like target in 2D.} We start with the example of a point-like target ($\epsilon=0$) on the surface of a 2D sphere. Note that, as opposed to  narrow-escape problems for one-state diffusion \cite{Schuss:2007},  the limit $\epsilon\to0$ is not singular and the MFPT at the target is finite, thanks to the one-dimensional surface diffusion. This MFPT satisfies the following backward equations \cite{Redner:2001}:
\begin{equation}
\label{t1}
D_1\Delta_{\partial S}t_1(\theta)+\lambda [t_2(R-a,\theta)-t_1(\theta)]=-1,
\end{equation}
 \begin{equation}
 \label{t2}
D_2\Delta _{S}t_2(r,\theta)=-1,
\end{equation}
where $t_1$ stands for the MFPT starting from the adsorbed state at a position defined on the surface by the polar angle $\theta$, and $t_2$ for the MFPT starting 
from the point $(r,\theta)$ in the bulk. Here, $\Delta_{\partial S}=\partial^2_{\theta}/R^2$ and $\Delta _{S}=\partial^2_{r}+\partial_r/r+\partial^2_{\theta}/r^2$. In Eqs(\ref{t1}),(\ref{t2}) the first term of the lhs accounts for the diffusion respectively on the surface and in the bulk, while
the second term of Eq.(\ref{t1}) describes desorption events. They have to be completed by boundary conditions:
 $t_2(R,\theta)=t_1(\theta)$,
 which describes the adsorption events and 
 $t_1(\theta=0)=0=t_1(\theta=2\pi)$,
which expresses that the target is an absorbing point in the problem.
As we proceed to show, these equations can be solved exactly. 

Considering $t_2$ as a source term in the Poisson type equation (\ref{t1}) with absorbing conditions at $\theta=0$ and $\theta=2\pi$, whose Green function is well known  \cite{Barton:1989}, $t_1$ writes:
\begin{eqnarray}
\label{t1integre}
t_1(\theta)&=&\frac{1}{\omega \sinh(2\pi \omega)}\int _0 ^{2\pi} \sinh(\omega \theta_<)\sinh (\omega(2\pi - \theta_>))\times\nonumber \\
&&\left[\frac{R^2}{D_1}+\frac{\lambda R^2}{D_1} t_2(R-a,\theta')\right]{\rm d} \theta',
\end{eqnarray}
where we have used the dimensionless variable $\omega\equiv R\sqrt{\lambda/D_1}$ and the notations $\theta_<=\min(\theta,\theta')$ and $\theta_>=\max(\theta,\theta')$.
On the other hand, Eq.(\ref{t2}) is easily shown to be satisfied by the following Fourier series:
\begin{equation}
\label{t2integre}
t_2(r,\theta)=\alpha_0-\frac{r^2}{4D_2}+\sum_{n=1}^\infty \alpha_n\left(\frac{r}{R}\right)^n \cos(n\theta),
\end{equation}
where the unknown coefficients $\alpha_n$ have to be calculated.
We aim at determining the reaction time, defined here as the MFPT at the target, for a molecule initially uniformly distributed on the circumference, {\it ie} $\langle t_1 \rangle \equiv \frac{1}{2\pi}\int_0^{2\pi } t_1(\theta){\rm d} \theta $.
Substituting Eqs(\ref{t1integre}),(\ref{t2integre}) in the boundary conditions leads after straightforward integrations to 
\begin{eqnarray}
\label{inter}
&&\frac{1}{\omega^2}\left(\alpha_0-\frac{R^2}{4D_2}+\sum_{n=1}^\infty \alpha_n  \cos(n\theta)\right)\cosh(2\pi \omega) =\nonumber\\
&&\!\!\!\!\!\!\!\!\!\!\!\frac{2\sinh(\pi \omega)}{\omega}\left(\frac{1}{\lambda}+\alpha_0-\frac{R^2x^2}{4D_2}\right)\left(\cosh(\pi\omega)-\cosh(\omega(\theta-\pi)\right)\nonumber\\
&+&\omega \sinh(2\pi\omega) \sum_{n=1}^\infty \alpha_n \frac{x^n}{\omega^2+n^2} \cos(n\theta)\nonumber\\
&-&2\omega{\sinh(\pi \omega)\cosh(\omega(\theta-\pi))}\sum_{n=1}^\infty \alpha_n \frac{x^n}{\omega^2+n^2},
\end{eqnarray}
where $x\equiv 1-a/R$.
The projection of  Eq(\ref{inter}) on the functions $\cos(n\theta)$ leads to an infinite hierarchy of equations for $\alpha_n$, $n\in \mathbb{N}$, which  can be decoupled.  It  finally yields an exact expression of the reaction time:
\begin{eqnarray}
\label{res}
\langle t_1 \rangle =\left[\frac{1}{\lambda}+\frac{R^2}{4D_2}(1-x^2)\right]
\left[\sum_{m=1}^\infty \frac{2\omega^2}{\omega^2\left(1-x^m\right)+m^2}\right].
\end{eqnarray}
Several comments are in order. (i)  As expected, both  limits $\lambda \to 0$ and $a\to 0$   of the reaction time are given by $\pi^2R^2/(3D_1)$ which corresponds to a pure $1D$ diffusion, with a molecule initially
uniformly distributed in $[0,2\pi R]$ with absorbing boundaries \cite{Redner:2001}. (ii)  The limit $a\to R$ of Eq.(\ref{res}) corresponds actually to rebinding positions uniformly distributed on the surface, and can therefore
be recovered from the  MF treatment of \cite{Obenichou:2008}, which ignores the correlations between starting and ending points of bulk excursions. Actually, it can be checked that this type of MF treatment, which
has been successfully used in the context of target search on DNA \cite{Obenichou:2008},  substantially underestimates the reaction time in   the physical regime $a\ll R$ in the present case of interfacial reactions, where correlations play a crucial role. Note that the MF approach of \cite{gleb} even misses the non-monotonicity with $\lambda$. (iii) Eq.(\ref{res}) has a clear physical interpretation: the reaction time is the product of the mean duration of each elementary step composed of one surface exploration and one excursion in  the bulk phase,  
by the mean number of excursions before reaction. This factorized structure is expected in the limit of large number of excursions since the durations of elementary steps are independent variables, all independent  of the number of excursions, except for the very last one. 

\begin{figure}[hbt]
\begin{centering}
\includegraphics[width=0.4\textwidth]{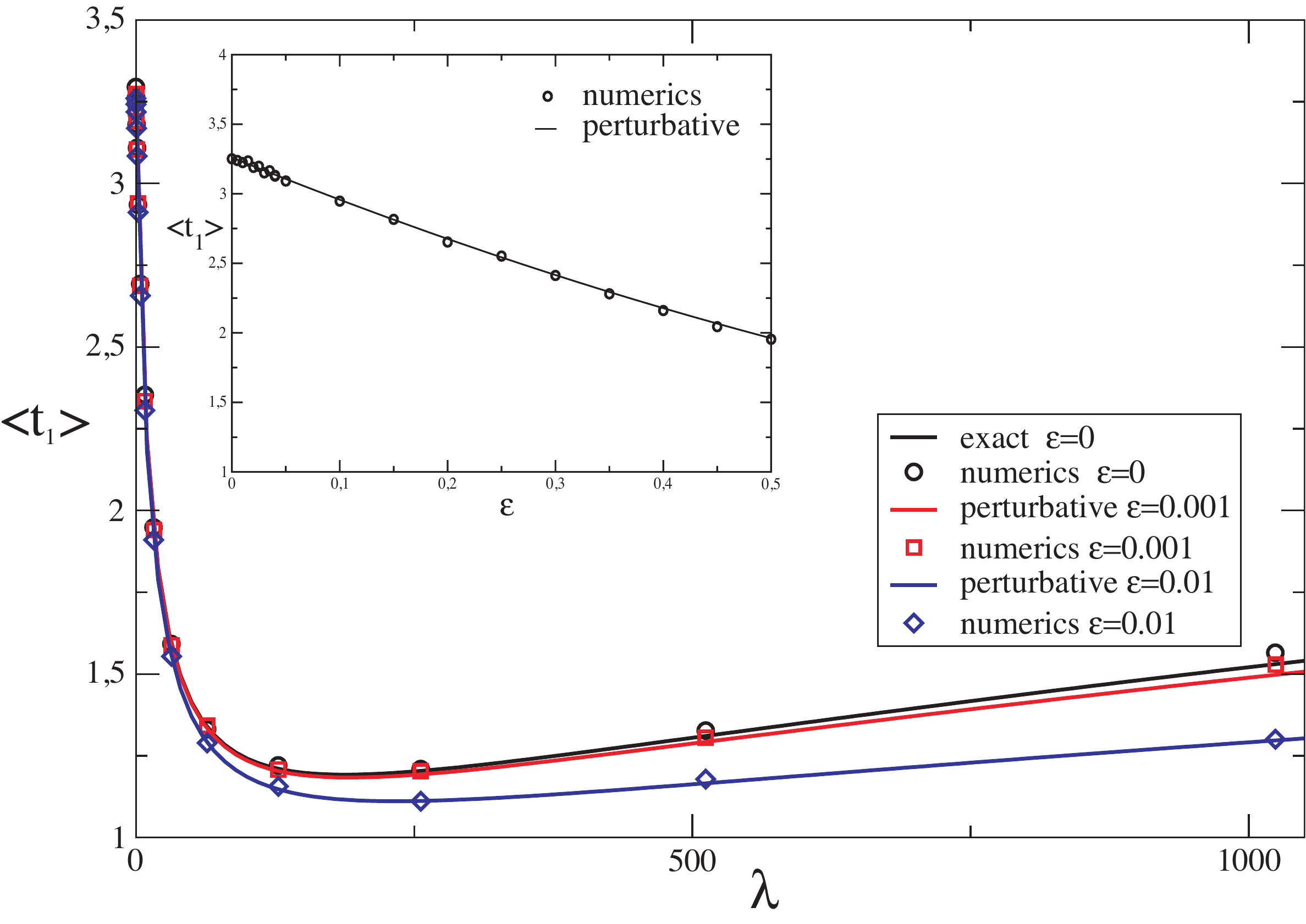}
\caption{\label{fig2}Mean reaction time as a function of the desorption rate in the 2D case, with $D_1=1$, $D_2=5$,  $a=0.1$ and $R=1$ (in arbitrary units): analytical perturbative expression Eq.(\ref{2Detenduebis}) vs Monte-Carlo numerical simulations for different target sizes $\epsilon$. Insert: mean reaction time as a function of the target size for a fixed desorption rate $\lambda=0.125$.}
\end{centering}
\end{figure}

Importantly, a first-order small $\lambda$ expansion of the reaction time allows one to show that the reaction time is a non-monotonic function of the desorption rate $\lambda$ if 
\begin{eqnarray}
\frac{D_1}{D_2}<\frac{24}{\pi^2(1-x^2)}\sum_{n=1}^\infty\frac{1}{n^4}(1-x^n),
\end{eqnarray}
which becomes $D_2/D_1>\pi^2/(12 \zeta(3))\approx 0.68 ...$ in the physically relevant limit $a/R \ll 1$ (ie $x\to1$), where $\zeta$ is the Riemann function. In other words,  even if $D_2$ is smaller than $D_1$, bulk excursions turn out to
speed up the reaction.
Focusing  on this limit $a/R \ll 1$ and $D_1\ll D_2$, a detailed analysis of the reaction time given explicitly by Eq.(\ref{res}) shows that the optimal $\lambda$ is given to leading order by 
 $\lambda^{{\rm opt}}\approx 2D_2\ln(2D_2/D_1)/(aR)$. In turn,  the gain, defined as the ratio of the reaction time in absence of bulk excursion over the optimal reaction time,  is found to be proportional to $D_2/D_1$, up to logarithmic corrections. This shows that for fast bulk diffusion,  reactivity can be   significantly  enhanced by surface-mediated diffusion by a proper tuning of the affinity of the reactants for the surface (see fig.\ref{fig2} for $\epsilon=0$). 

{\it Extended target in 2D. }The previous analysis can be generalized to the important  case of an extended target zone (of angular extension $2\epsilon$), especially relevant to the case of escape problems \cite{Schuss:2007}. The calculation leads  
in this case to an infinite hierarchy of coupled equations for $\alpha_n, \;n\in\mathbb N$, in contrast to the point-like target case. 
This additional    complexity results from the fact that the target can now be reached not only from the surface, but also  directly from the bulk. The first terms of an exact perturbative expansion in $\epsilon$ can  be obtained and read:
\begin{eqnarray}
\label{2Detenduebis}
&&\langle t_1(\epsilon) \rangle =\langle t_1 (\epsilon=0)\rangle+\omega^2\left[\frac{1}{\lambda}+\frac{R^2}{4D_2}(1-x^2)\right]\times\nonumber\\
&&\!\!\!\!\!\times\!\!\left[-\pi \epsilon + \left(1+\sum_{m=1}^\infty \frac{2\omega^2(1-x^m)}{\omega^2\left(1-x^m\right)+m^2}\right)\epsilon^2 + ... \right]
\end{eqnarray}
Note that the coefficients of $\epsilon^k$ of this expansion diverge with $\omega$, so that in practice,  the smaller $\omega$, the wider  the range of applicability in $\epsilon$. Fig.\ref{fig2} 
shows an excellent quantitative agreement even for rather large values of $\epsilon$.

In addition, the benefit of bulk excursions can still be analyzed with $\epsilon\neq0$.  A small $\lambda$ expansion of the reaction time can  be worked out, and shows that bulk excursions reduce the reaction time provided that:
\begin{eqnarray}
\label{2Detendue}
\frac{D_1}{D_2}<\frac{ \sum_{n=1}^\infty \frac{24}{n^6}\left(1-x^n\right)\left(n(\pi-\epsilon)\cos(n\epsilon)+\sin(n\epsilon)\right)^2}{\pi (1-x^2)(\pi-\epsilon)^3}.\nonumber\\
\end{eqnarray}

\begin{figure}[hbt]
\begin{centering}
\includegraphics[width=0.4\textwidth]{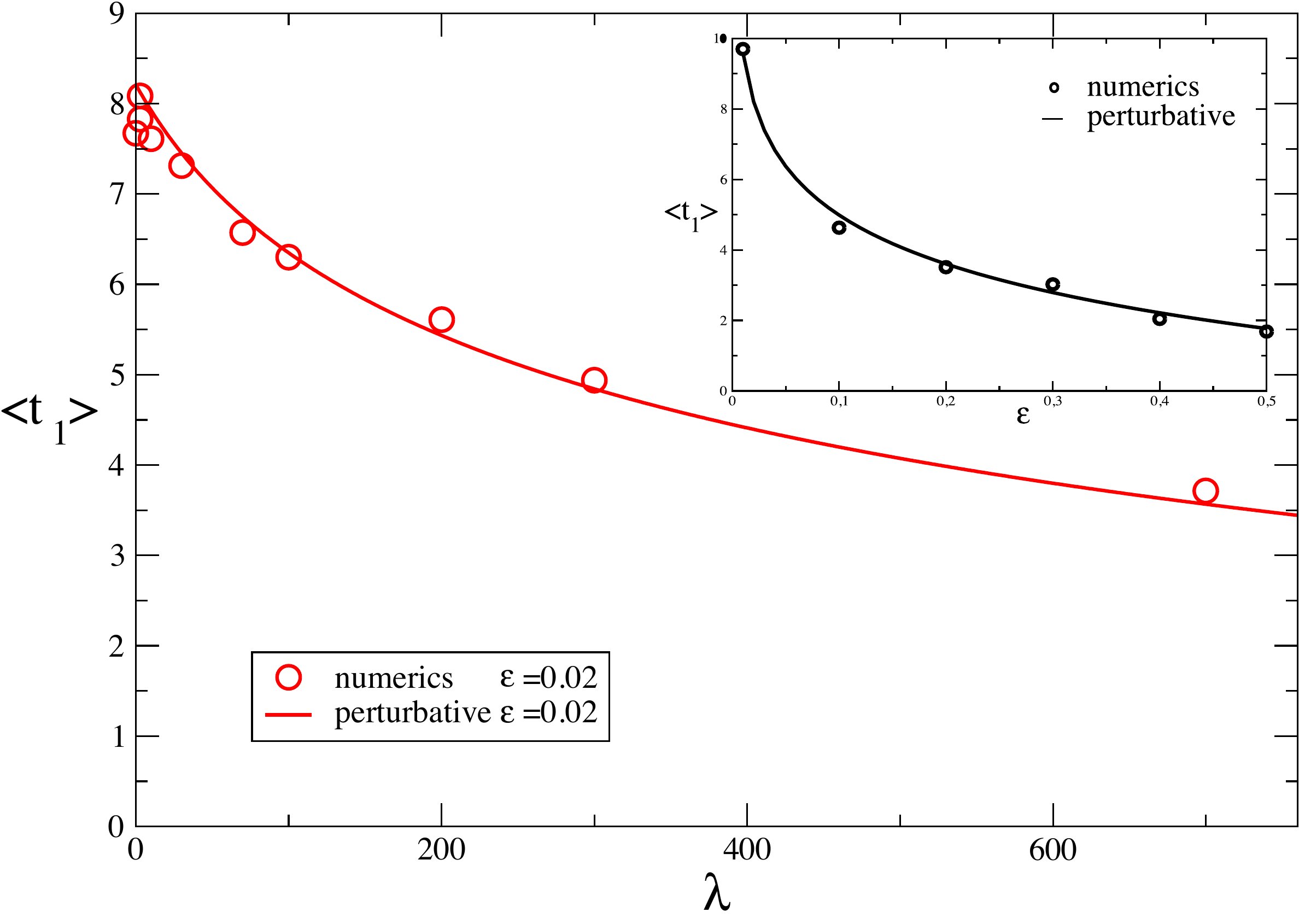}
\caption{\label{fig3}Mean reaction time as a function of the desorption rate in the 3D case with $D_1=1$, $D_2=50$,  $a=0.02$ and $R=1$ (in arbitrary units): analytical perturbative expression Eq.(\ref{t3D}) vs Monte-Carlo simulations for a target of size $\epsilon=0.02$. Insert: mean reaction time as a function of the target size for a fixed desorption rate $\lambda=0.125$.}
\end{centering}
\end{figure}


 {\it 3D case. }The previous analysis can also be adapted to the case of a confining 3D sphere, relevant to many situations such as reactions in micellar or vesicular systems \cite{Sano:1981,Schuss:2007}. Eqs(\ref{t1})-(\ref{t2}) are still valid 
 in this case with $\Delta_{\partial S}=\partial_\theta(\sin \theta \partial_\theta)/(R^2\sin \theta)$ and $\Delta_{S}$ is the usual 3D Laplacian. Along the same lines (technical details are given in Supplementary Information), the first  terms of an exact  small $\epsilon$ expansion can be obtained  and read :
 \begin{eqnarray}
 \label{t3D}
&&\langle t_1(\epsilon) \rangle = \omega^2\left(\frac{1}{\lambda}+\frac{R^2}{6D_2}(1-x^2)\right)\times\\
&&\!\!\!\!\!\!\!\!\!\!\times\!\!\left[2\ln\left(\frac{2}{\epsilon}\right)-1-\sum_{n=1}^\infty \frac{2n+1}{n(n+1)}\frac{\omega^2(1-x^n)}{n(n+1)+\omega^2(1-x^n)}\right]+...\nonumber
\end{eqnarray}
Similarly to the 2D case, one can show that bulk excursions are beneficial under the condition that now reads
\begin{eqnarray}
\frac{D_1}{D_2}<\frac{\sum_{n=1}^\infty (1-x^n)\frac{3(2n+1)}{n^2(n+1)^4}b_n}{2(1-x^2)\left(\ln(2/(1-\cos\epsilon))-(1+\cos\epsilon)/2\right)}
\end{eqnarray}
where $b_n\equiv((n\cos\epsilon+n+1)P_n(\cos\epsilon)+P_{n-1}(\cos\epsilon))^2$ and $P_n$ stand for Legendre polynomials, and that the reaction time can be minimized.  Fig.\ref{fig3} shows a good quantitative agreement of the expansion Eq.(\ref{t3D})  with Monte-Carlo simulations and confirms that the reaction time can be decreased by bulk excursions, as in the 2D situation.

 \begin{figure}[hbt]
\begin{centering}
\includegraphics[width=0.4\textwidth]{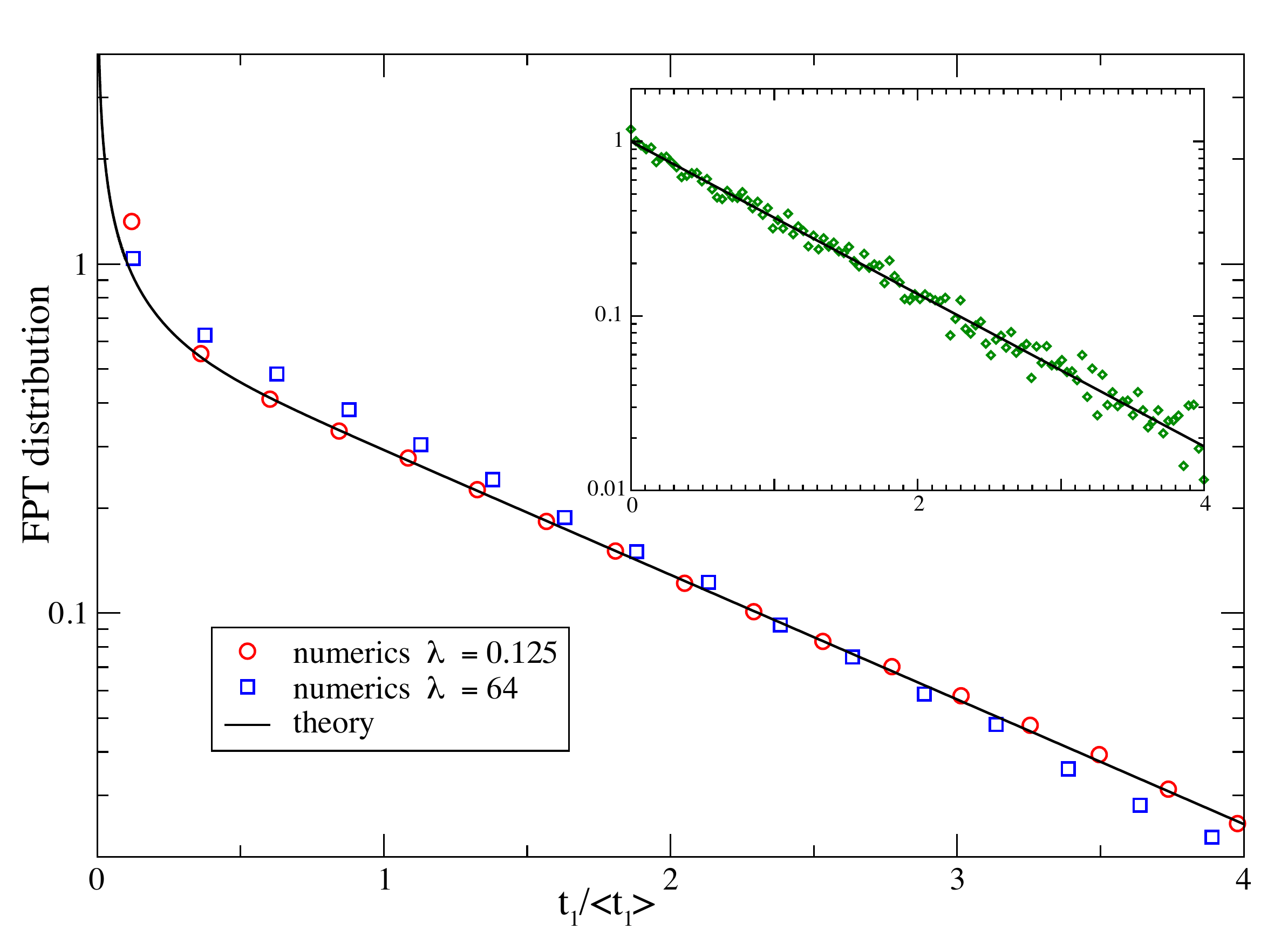}
\caption{\label{fig4} FPT distribution of the reduced variable $t_1/\langle t_1 \rangle$ : theory (plain line) vs simulations (symbols).  In the 2D case, $D_1=1$, $D_2=10$,  $a=0.1$, $\epsilon=0.01$  and $R=1$ (in arbitrary units). Insert: 3D case, with $D_1=1$, $D_2=50$,  $a=0.02$, 
$\lambda=70$, $\epsilon=0.02$  and $R=1$.}
\end{centering}
\end{figure}

 {\it Entire FPT distribution.} Last,  using the results given recently in \cite{BenichouO.:2010,bis}, we stress that an estimate of the {\it entire} FPT distribution of the reduced variable $t_1/\langle t_1 \rangle$ can be  inferred in the large size limit $R\to\infty$  from  the knowledge of the MFPT calculated above.    While an exponential form is predicted in the 3D case (since the exploration of the boundary is  marginally compact), the FPT distribution in the 2D case involves several time scales, and is given by a sum of exponentials (see Eq.(5) of \cite{bis}, case of compact  exploration). Fig. \ref{fig4} shows a good agreement of this theoretical prediction with numerical simulations in both cases.

In conclusion, we have presented an exact calculation of the mean first-passage time to a small target on the surface of a 2D and 3D spherical domain, for a molecule performing surface-mediated diffusion. This minimal model, which explicitly  
takes into account the combination of surface and bulk diffusion, shows that bulk excursions can speed up the reaction as they reduce oversampling of the boundary and underlines the importance of correlations induced by the coupling of the switching dynamics to the geometry of the confinement.
In the context of interfacial systems in confinement, our results show that the reaction time can be minimized as a function of the desorption rate from the surface, which puts forward a general mechanism of enhancement and 
regulation of chemical reactivity.


\end{document}